\documentstyle[prl,aps,multicol]{revtex}

\input{epsf}

\begin{document}
\draft
\title{Stochastic wave function approach to generalized master equations}

\author{Heinz--Peter Breuer, Bernd Kappler and Francesco Petruccione}
\address{
Albert-Ludwigs-Universit\"at, Fakult\"at f\"ur Physik, \\
Hermann-Herder Stra{\ss}e 3, D--79104 Freiburg im Breisgau,
Federal Republic of Germany}
\date{\today}
\maketitle

%------------------------------------------------------------
% Abstract 
%------------------------------------------------------------
\begin{abstract}
A generalization of the stochastic wave function method is presented
which allows the unravelling of arbitrary linear quantum master
equations which are not necessarily in Lindblad form and, moreover, the
explicit treatment of memory effects by employing the
time-convolutionless projection operator technique. The crucial point
of this construction is the description of the open system in a doubled
Hilbert space, which has already been successfully used for the
computation of multitime correlation functions. 
\end{abstract}

%----------------------------------------------------------
\begin{multicols}{2}
\narrowtext

%----------------------------------------------------------
%   Introduction
%----------------------------------------------------------
\section{Introduction}
\label{sec:Intro}

Usually, the state of an open quantum system is described by a reduced
density matrix $\rho(t)$ which is a positive operator on the
Hilbert space ${\cal H}_{\rm S}$ of the system. On the other hand,
within the stochastic wave function method the state of the open
system is described by an ensemble of pure, normalized states $\psi(t)$
the covariance matrix of which equals the reduced density matrix 
\cite{Carmichael,MolmerPRL68,Gisin:92,GardinerPRA46,BP:QS4},
\begin{equation}
  \label{eq:P_rho}
  \rho(t)=\int D\psi D\psi^*|\psi\rangle\langle\psi|P[\psi,t].
\end{equation}
In Eq.~(\ref{eq:P_rho}) the integral extends over the Hilbert space of
the system, $D\psi D\psi^*$ denotes the Hilbert space volume element,
and $P[\psi,t]$ is the time-dependent probability density of finding
the state of the system in the volume element $D\psi D\psi^*$ near
$\psi$ \cite{BP:QS4}. This formulation has essentially two advantages
compared to the conventional description: First, this approach allows
the investigation of the dynamics of an individual quantum system
which is continuously observed by some measurement device
\cite{WisemanPRA93,BP:QS8}, whereas the reduced density matrix can
only describe the state of an ensembles of quantum systems.  Second,
from a computational point of view, the numerical integration of the
quantum master equation can become rather expensive for large systems,
since the reduced density matrix has $N^2$ degrees of freedom, where
$N$ is the dimension of the system's Hilbert space. In contrast, a
stochastic wave function has only $N$ components, which can
significantly reduce the computational expense
\cite{BP:QS11}. Moreover, algorithms which are based on stochastic
simulations can easily be implemented on parallel computers.

The dynamics of the stochastic wave function is governed by a stochastic
evolution equation, and the construction of this evolution equation
within the Born-Markov approximation is well understood. However, in
some situations non-Markovian effects can significantly alter the
reduced system dynamics. In this article we will present a scheme
which allows a systematic incorporation of memory effects into
the stochastic wave function method. To this end, we make use of an
expansion scheme which is known from the theory of
non-equilibrium statistical mechanics -- the time-convolutionless
projection operator technique \cite{ShibataZPhysB,ShibataJStat}. 

This article is organized as follows. In Sec.~\ref{sec:stqme} we
discuss the unravelling of quantum master equations by stochastic wave
functions. This concept is well known for quantum master equations
which are in Lindblad form \cite{Alicki} and we briefly summarize the major
results in Sec.~\ref{sec:stlind}. In Sec.~\ref{sec:SGQME} we
generalize this concept to the treatment of arbitrary linear quantum
master equations. Using this result, we present in Sec.~\ref{sec:TCL}
a general framework which allows an explicit treatment of memory
effects within the stochastic wave function method. This concept is
then illustrated by means of an exactly solvable model in
Sec.~\ref{sec:Ex} -- the damped Jaynes-Cummings model.

%----------------------------------------------------------
%   Stochastic simulation of quantum master equations 
%----------------------------------------------------------
\section{Stochastic simulation of quantum master equations}
\label{sec:stqme}

%----------------------------------------------------------
%   Quantum master equations in Lindblad form
%----------------------------------------------------------
\subsection{Quantum master equations in Lindblad form}
\label{sec:stlind}

In the Markovian regime, the time evolution of the reduced density
matrix $\rho(t)$ is governed by the quantum master equation in
Lindblad form
\begin{eqnarray}
  \label{eq:QME}
  \lefteqn{\frac{\partial}{\partial
  t}\rho(t)=-i\left[H_{\rm S}+\frac{1}{2}\sum_iS_i(t)L_i^\dagger L_i,\rho(t)\right]}\\
  &+&\sum_i\gamma_i(t)\left\{-\frac{1}{2}L_i^\dagger L_i\rho(t)
  -\frac{1}{2}\rho(t)L_i^\dagger L_i+L_i\rho(t)L_i^\dagger\right\}, \nonumber
\end{eqnarray}
where $H_{\rm S}$ is the Hamiltoninan of the system, the
time-dependent coefficients $S_i(t)$ describe an energy shift induced
by the coupling to the environment, namely the Lamb and Stark shifts,
and the positive rates $\gamma_i(t)$ model the dissipative coupling to
the $i-$th decay channel. This evolution equation is either obtained
by a phenomenological ansatz or through a derivation which is based on
a microscopic model of the system-reservoir interaction.

Using similar techniques, one can also obtain a Markovian time
evolution equation for the stochastic state vector $\psi$. There are
several phenomenological approaches which simply construct a
stochastic evolution equation in such a way, that the equation of
motion of the covariance matrix is the quantum master equation
\cite{MolmerPRL68,Gisin:92,GardinerPRA46}. This procedure is often
called {\em unravelling of the quantum master equation}
\cite{Carmichael}. Other approaches are based on a continuous
observation of the system under consideration by some measurement
device, for example a photon detector, and employ the basic
measurement postulates for the description of the dynamics of an
individual quantum system \cite{WisemanPRA93,BP:QS8}. Finally, similar
to the derivation of the quantum master equation, it is also possible
to obtain the stochastic time evolution directly from an underlying
microscopic model by an explicit derivation of the differential
Chapman Kolmogorov equation for the probability density $P[\psi,t]$
\cite{BP:QS4}.

A particular example of such a stochastic evolution equation which
arises in the above approaches is the stochastic differential
equation
\begin{equation}
  \label{eq:SDE}
  d\psi(t)=-iG(\psi,t) dt+
  \sum_i\left(\frac{L_i\psi(t)}{\|L_i\psi(t)\|}-\psi(t)\right)dN_i(t),
\end{equation}
where the $dN_i(t)$ are the differentials of independent Poisson
process $N_i(t)$ with mean $\langle
dN_i(t)\rangle=\gamma_i(t)\|L_i\psi(t)\|^2dt$ and
\begin{eqnarray}
  \label{eq:G_eq}
  G(\psi,t)&=&H(t)\psi+\frac{1}{2}\sum_iS_i(t) L_i^\dagger L_i\psi\nonumber\\
  &&-\frac{i}{2}\sum_i\gamma_i(t)\left(L_i^\dagger L_i -\|L_i\psi\|^2 
  \right)\psi.
\end{eqnarray}
This particular equation of motion describes the time
evolution of a piecewise deterministic process. The differential of
the Poisson process $dN_i(t)$ can either take the value $0$ or $1$. If
$dN_i(t)=0$, then the system evolves continuously according to the
nonlinear Schr\"odinger-type equation 
\begin{equation}
  \label{eq:cont}
  i\frac{\partial}{\partial t}\psi(t)=G(\psi,t),
\end{equation}
whereas, if $dN_i(t)=1$ for some $i$, then the system undergoes an
instantaneous, discontinuous transition of the form
\begin{equation}
  \label{eq:trans}
  \psi(t)\longrightarrow\frac{L_i\psi(t)}{\|L_i\psi(t)\|}.
\end{equation}
Note that the generator $G(\psi,t)$ of the continuous time evolution 
is non-Hermitian and hence the propagator of $\psi(t)$ is non-unitary.
However, due to the nonlinearity of the generator, the norm of
$\psi(t)$ is preserved in time.

Using the standard Ito calculus for the differentials $dN_i(t)$ of a
Poisson process, i.~e., $dN_i(t)dN_j(t)=\delta_{ij} dN_i(t)$, it is
easy to check, that the equation of motion of the covariance matrix of
$\psi(t)$ equals the usual Markovian quantum master equation for the
reduced density matrix in Lindblad form. Thus, both descriptions yield
the same equations of motion for the expectation values of system
observables. Finally, we want to remark that it is also possible to extend the
stochastic wave functions method to the calculation of arbitrary
matrix elements of system operators and hence to the determination of
multitime correlation functions \cite{GardinerPRA46,BP:QS14}. 

%----------------------------------------------------------
%   General quantum master equations
%----------------------------------------------------------
\subsection{General quantum master equations}
\label{sec:SGQME}

In this section we present a generalization of the stochastic wave
function method to quantum master equations which are not in Lindblad
form  (in Sec.~\ref{sec:TCL} we will also encounter this type of
evolution equations). To be more specific, we consider an equation of
motion for the reduced density matrix of the form
\begin{equation}
  \label{eq:rho_mo}
  \frac{\partial}{\partial t}\rho(t)=A(t)\rho(t)+\rho(t)B^\dagger(t)+\sum_i
  C_i(t)\rho(t)D_i^\dagger(t),
\end{equation}
with some arbitrary time-dependent linear operators $A(t)$, $B(t)$,
$C_i(t)$, and $D_i(t)$. This form represents the most general linear
equation of motion for $\rho(t)$, which is local in time, i.~e., an
equation of motion where $\dot\rho(t)$ only depends on $\rho(t)$. In
order to find an unravelling of this equation of motion we follow a
strategy, which has already been successfully applied to the
calculation of multitime correlation functions \cite{BP:QS14}: We
describe the state of the open system by a pair of stochastic state
vectors $\theta=(\phi,\psi)^T$ which is an element of the doubled
Hilbert space $\widetilde{\cal H}={\cal H}\oplus{\cal H}$, in such a
way, that
\begin{equation}
  \label{eq:rho_dbl_P}
  \rho(t)=\int D\theta D\theta^* |\phi\rangle\langle\psi|\widetilde
  P[\theta,t], 
\end{equation}
where the integral extends over the doubled Hilbert space
$\widetilde{\cal H}$, and $\widetilde P[\theta,t]$ is the probability
density of finding the system in the ``state'' $\theta$ at time $t$.
Furthermore, we define the operators $F(t)$ and $J_i(t)$ as 
\begin{equation}
  \label{eq:AB}
   F(t)=
  \left(\begin{array}{cc}
  A(t)&0\\
  0&B(t)\end{array}\right),\quad
   J_i(t)=
  \left(\begin{array}{cc}
  C_i(t)&0\\
  0&D_i(t)\end{array}\right).
\end{equation}
An unravelling of the quantum master equation (\ref{eq:rho_mo}) by a
stochastic wave function $\theta(t)$ can be obtained using the
stochastic differential equation 
\begin{eqnarray}
  \label{eq:gen_un}
  d\theta(t)&=&-iG(\theta,t)dt\\
  &&+\sum_i\left(\frac{\|\theta(t)\|}{\left\|
  J_i(t)\theta(t)\right\|} 
  J_i(t)\theta(t)-\theta(t)\right)dN_i(t) \nonumber,
\end{eqnarray}
where $dN_i(t)$ is the differential of a Poisson process with mean
\begin{equation}
  \label{eq:dbl_mean}
  \langle dN_i(t)\rangle=\frac{\left\| J_i(t)\theta(t)\right\|^2}
  {\|\theta(t)\|^2}dt,
\end{equation}
and 
\begin{equation}
  \label{eq:G_dbl}
   G(\theta,t)=i\left( F(t)+\frac{1}{2}\sum_i 
  \frac{\left\| J_i(t)\theta(t)\right\|^2}{\|\theta(t)\|^2}\right)
  \theta(t).
\end{equation}
Again, the stochastic differential equation (\ref{eq:gen_un})
describes a piecewise deterministic jump process, where $G(\theta,t)$
is the generator of the continuous time evolution and the operators
$J_i(t)$ lead to discontinuous instantaneous transitions. In order to
show that the stochastic differential equation (\ref{eq:gen_un}) leads
to the correct equation of motion for $\rho(t)$ one can rewrite
Eq.~(\ref{eq:gen_un}) as a system of coupled stochastic differential
equations for $\phi$ and $\psi$, and compute the mean of the differential 
\begin{equation}
  \label{eq:diff}
  d\left(|\phi\rangle\langle\psi|\right)=|d\phi\rangle\langle\psi|
  +|\phi\rangle\langle d\psi|+|d\phi\rangle\langle d\psi|
\end{equation}
using the Ito calculus, which justifies our ansatz. It is important
to note that this unravelling also contains the unravelling presented in
Sec.~\ref{sec:stlind}. If the equation of motion for the reduced
density matrix is in Lindblad form, and $\phi(0)=\psi(0)$ then it is
easy to see that $\phi(t)$ and $\psi(t)$ are identical for all $t$ and
the stochastic differential equation (\ref{eq:gen_un}) reduces to
Eq.~(\ref{eq:SDE}) with the choice
\begin{equation}
  \label{eq:A_B_eq}
  A(t)=B(t)=
  -iH_{\rm S}
  -\frac{1}{2}\sum_k\left[\gamma_k(t) +iS_k(t)\right] L_k^\dagger L_k
\end{equation}
and
\begin{equation}
  \label{eq:C_D}
  C_i(t)=D_i(t)=\sqrt{\gamma(t)}L_i.
\end{equation}

%----------------------------------------------------------
%   Non-Markovian stochastic wave function method
%----------------------------------------------------------
\section{Non-Markovian stochastic wave function method}
\label{sec:TCL}

In this section we present a general scheme which allows a systematic,
perturbative treatment of memory effects within the stochastic wave
function method. This approach is based on the time-convolutionless
projection operator technique \cite{ShibataZPhysB,ShibataJStat}, which
is related to the Nakajima-Zwanzig projection operator technique
\cite{Nakajima,Zwanzig,Resibois}.

As a microscopic model, we consider a system which is coupled to an
environment. The Hamiltonian of the total system is given by
\begin{equation}
  \label{eq:H_tot}
  H=H_{0}+\alpha H_{\rm I},
\end{equation}
where $H_0$ describes the free evolution of the system and the
reservoir and $H_{\rm I}$ their interaction. The parameter $\alpha$
denotes a dimensionless expansion parameter. The state of the total
system is described by the interaction picture density matrix $W(t)$
which is a solution of the Liouville-von Neumann equation
\begin{equation}
  \label{eq:LvN}
  \frac{\partial}{\partial t}W(t)=-i\alpha[H_{\rm I}(t),W(t)]\equiv
  \alpha L(t)W(t),
\end{equation}
where the interaction Hamiltonian in the interaction picture is
defined as $H_{\rm I}(t) = \exp(iH_{0}t) H_{\rm I} \exp(-iH_{0}t)$.
Since we are interested in the dynamics of the reduced system, we
define a projector ${\cal P}$ as
\begin{equation}
  \label{eq:projector}
  {\cal P}W(t)=\mbox{Tr}_{\rm R}\left\{W(t)\right\}\otimes \rho_{\rm
  R}\equiv\rho(t)\otimes\rho_{\rm R},
\end{equation}
where $\rho_{\rm R}$ is a stationary state of the environment, and a
projector ${\cal Q}= 1-{\cal P}$.  A quasi-closed equation of motion
for ${\cal P}W(t)$ can be obtained by using the Nakajima-Zwanzig
projection operator technique \cite{Nakajima,Zwanzig,Resibois},
namely
\begin{equation}
  \label{eq:GME}
  \frac{\partial}{\partial t}{\cal P}W(t)=\int_0^tds \widetilde
  K(t,s){\cal P}W(s),
\end{equation}
where the memory kernel $\widetilde K(t,s)$ is defined as
\begin{equation}
  \label{eq:K_tild}
  \widetilde K(t,s)=\alpha^2{\cal P}L(t){\cal G}(t,s)L(s),
\end{equation}
with the propagator
\begin{equation}
  \label{eq:prop}
  {\cal G}(t,s)={\rm T}_\leftarrow\exp\left[\alpha\int_s^tds'  
  {\cal Q}L(s')\right]
\end{equation}
and ${\rm T}_\leftarrow$ indicates the chronological time ordering. In
obtaining Eq.~(\ref{eq:GME}) we have assumed that $\mbox{Tr}_{\rm
R}\{H_{\rm I}^{2k+1}\rho_{\rm R}\}=0$, and that the system and the
reservoir are uncorrelated initially, i.~e.,
$W(0)=\rho(0)\otimes\rho_{\rm R}$.  To eliminate the time-convolution
\cite{ShibataZPhysB,ShibataJStat} in Eq.~(\ref{eq:GME}) we replace
$W(s)$ by the expression
\begin{equation}
  \label{eq:W_back}
  W(s)= G(t,s)({\cal P}+{\cal Q})W(t),
\end{equation}
where the backward propagator $G(t,s)$ is defined as 
\begin{equation}
  \label{eq:back_prop}
  G(t,s)={\rm T}_\rightarrow\exp\left[-\alpha\int_s^tds'L(s')\right]
\end{equation}
and ${\rm T}_\rightarrow$ indicates the anti-chronological time
ordering. This leads to the time-convolutionless equation of motion
\begin{equation}
  \label{eq:TCL}
  \frac{\partial}{\partial t}{\cal P}W(t)=K(t){\cal P}W(t),
\end{equation}
where the generator $K(t)$ is defined as 
\begin{equation}
  \label{eq:K}
  K(t)=\alpha{\cal P}L(t)[1-\Sigma(t)]^{-1}{\cal P}
\end{equation}
and 
\begin{equation}
  \label{eq:sigma}
  \Sigma(t)=\alpha\int_0^tds{\cal G}(t,s)L(s){\cal P}G(t,s).
\end{equation}
If the operator $(1-\Sigma(t))^{-1}$ can be expanded in a geometric
series, which is possible if the coupling between the system
and the reservoir is not too strong, then we can rewrite
the generator $K(t)$ as
\begin{equation}
  \label{eq:gen2}
  K(t)=\alpha\sum_{n=0}^\infty{\cal P}L(t)\left(\Sigma(t)\right)^n{\cal P}
\end{equation}
and obtain a perturbative expansion in the form 
\begin{equation}
  \label{eq:K_pert}
  K(t)=\sum_{n=0}^\infty \alpha^{2n}K_{2n}(t).
\end{equation}
Note that all terms containing odd orders of the coupling constants
vanish, since by definition of ${\cal P}$ and $L(t)$ we have ${\cal
P}L(t_1)\cdots L(t_{2k+1}){\cal P}=0$.  The explicit expressions
for the second and fourth order contribution are
\begin{equation}
  \label{eq:K_2}
  K_2(t)= \int_0^tdt_1{\cal P}L(t)L(t_1){\cal P},
\end{equation}
and 
\begin{eqnarray}
  \label{eq:K_4}
 \lefteqn{ K_4(t)  = \int_0^tdt_1\int_0^{t_1}dt_2\int_0^{t_2}
  dt_3}\\
  &&\times\Big[{\cal  P}L(t)L(t_1)L(t_2)L(t_3){\cal P}
  -{\cal  P}L(t)L(t_1){\cal  P}L(t_2)L(t_3){\cal P}\nonumber\\
  &&-{\cal  P}L(t)L(t_2){\cal  P}L(t_1)L(t_3){\cal P}
  -{\cal  P}L(t)L(t_3){\cal  P}L(t_1)L(t_2){\cal P}\Big].\nonumber
\end{eqnarray}
The higher order contributions can be obtained in a systematic way
by a slight modification of van Kampen's cumulant expansion \cite{kampenkumm2}
(see also \cite{ShibataZPhysB}).

The time-convolutionless quantum master equation (\ref{eq:TCL}) allows
us to use the stochastic wave function method for the description of
the dynamics of the open system. To this end, we note that the
equation of motion for the reduced density matrix $\rho(t)$ which
results from either using Eq.~(\ref{eq:TCL}) directly or any
perturbative approximation of this equation, is linear in $\rho(t)$
and local in time. Hence, we can write it in the form of
Eq.~(\ref{eq:rho_mo}) where, of course, the operators $A(t)$, $B(t)$,
$C_i(t)$, and $D_i(t)$ depend on the interaction Hamiltonian $H_{\rm
I}$, and apply the unravelling of this equation of motion described in
Sec.~\ref{sec:SGQME}. This leads to a stochastic wave function
description of the open system which can be formulated, at least in
principal, to any desired order in the coupling, and which is hence
generally applicable to any open quantum system.

%----------------------------------------------------------
%   Example
%----------------------------------------------------------
\section{Example}
\label{sec:Ex}

As a specific example for the general concept presented in
Secs.~\ref{sec:stqme} and \ref{sec:TCL} we consider the spontaneous
decay of a two-level system coupled to the electromagnetic field which
is initially in the vacuum state within the rotating wave
approximation. The Hamiltonian of the total system is given by 
\begin{eqnarray}
  \label{eq:H_0}
  H_{\rm 0}&=&\omega_{\rm S}\sigma^+\sigma^-+\sum_k\omega_k b_k^\dagger b_k,\\
  \label{eq:H_I}
  H_{\rm I}&=&\sigma^+\otimes B+\sigma^-\otimes B^\dagger \mbox{ with
  } B=\sum_k g_k b_k,
\end{eqnarray}
where $\omega_{\rm S}$ and $\omega_k$ denote the eigenfrequencies of
the system and reservoir, respectively, and the $g_k$ are real
coupling constants. As usual, $\sigma^\pm$ denote the pseudospin
operators, and the $b_k$ are the annihilation operators for the field
mode $k$. Inserting the above definitions into the expressions for
$K_2(t)$ and $K_4(t)$, Eqs.~(\ref{eq:K_2}) and (\ref{eq:K_4}), we
obtain an equation of motion for the reduced density matrix $\rho(t)$, the
time-convolutionless quantum master equation
\begin{eqnarray}
  \label{eq:TCL-app}
\lefteqn{\frac{\partial}{\partial t}\rho(t)=-\frac{i}{2}S^{(4)}(t)
  [\sigma^+\sigma^-,\rho(t)]}\\
  &+&\gamma^{(4)}(t)\left\{-\frac{1}{2}\sigma^+\sigma^-\rho(t)
  -\frac{1}{2}\rho(t)\sigma^+\sigma^- +\sigma^-\rho(t)\sigma^+
  \right\}, \nonumber
\end{eqnarray}
which is in Lindblad form with time-dependent coefficients. The
coefficients to fourth order are given by
\begin{eqnarray}
  \label{eq:Lamb}
 \lefteqn{ S^{(4)}(t)=\int_0^tdt_1\Psi(t-t_1)+\frac{1}{2}
  \int_0^tdt_1\int_0^{t_1}dt_2\int_0^{t_2}dt_3 }\nonumber\\
  &\times&\Big[\Psi(t-t_2)\Phi(t_1-t_3)+\Phi(t-t_2)\Psi(t_1-t_3)\nonumber\\
  &+&\Psi(t-t_3)\Phi(t_1-t_2)+\Phi(t-t_3)\Psi(t_1-t_2)\Big]
\end{eqnarray}
and
\begin{eqnarray}
  \label{eq:rate}
  \lefteqn{\gamma^{(4)}(t)=\int_0^tdt_1\Phi(t-t_1)+\frac{1}{2}\int_0^tdt_1\int_0^{t_1}
  dt_2\int_0^{t_2} dt_3}\nonumber\\ 
  &\times&\Big[\Psi(t-t_2)\Psi(t_1-t_3)-\Phi(t-t_2)\Phi(t_1-t_3)\nonumber\\
  &+&\Psi(t-t_3)\Psi(t_1-t_2)-\Phi(t-t_3)\Phi(t_1-t_2)\Big].
\end{eqnarray}
The real functions $\Phi(t)$ and $\Psi(t)$ are related to the
reservoir correlation functions through
\begin{eqnarray}
  \label{eq:phi_psi}
   \Phi(t)+i\Psi(t)&=&2 \mbox{Tr}_{\rm R}\left\{B(t)B^\dagger
   \rho_{\rm R}\right\}e^{i\omega_{\rm S}t}\nonumber\\ 
   &=&2\int d\omega J(\omega)e^{i(\omega_{\rm S}-\omega)t},
\end{eqnarray}
where $B(t)=\exp(iH_0t)B\exp(-iH_0t)$, and we have performed the
continuum limit. The function $J(\omega)$ is the spectral density times the
strength of the coupling at the frequency $\omega$. This function
determines the statistical properties of the system dynamics. 

For this particular model an exact equation of motion for the reduced
density matrix can be obtained in the following way: First we define
the three states \cite{GarrawayPRA55}
\begin{eqnarray}
  \label{eq:expand}
  \psi_0&=&|0\rangle_{\rm S}\otimes|0\rangle_{\rm R}\nonumber\\
  \psi_1&=&|1\rangle_{\rm S}\otimes|0\rangle_{\rm R}\nonumber\\
  \psi_k&=&|0\rangle_{\rm S}\otimes|k\rangle_{\rm R}
\end{eqnarray}
where $|0\rangle_{\rm S}$ and $|1\rangle_{\rm S}$ indicate the ground
and excited state of the system, respectively, the state
$|0\rangle_{\rm R}$ denotes the vacuum state of the reservoir, and
$|k\rangle_{\rm R}=b_k^\dagger|0\rangle_{\rm R}$ denotes the state
with one photon in mode $k$. If an initial pure state $\phi(0)$
can be expanded in terms of these states, then the state at time $t$
has the form \cite{GarrawayPRA55}
\begin{equation}
  \label{eq:phi_exp}
  \phi(t)=c_0\psi_0+c_1(t)\psi_1+\sum_k c_k(t)\psi_k,
\end{equation}
with some probability amplitudes $c_0$, $c_1(t)$, and $c_k(t)$
and, hence, the reduced density matrix is given by 
\begin{equation}
  \label{eq:rho_ex}
  \rho(t)=\left(\begin{array}{cc}
  |c_1(t)|^2& c_1(t) c_0^*\\
  c_1^*(t)c_0& |c_0|^2+\sum_k|c_k|^2
  \end{array}\right).
\end{equation}
Differentiating $\rho(t)$ with respect to time leads
to a quantum master equation in the Lindblad form
\begin{eqnarray}
  \label{eq:TCL-QME}
  \frac{\partial}{\partial t}\rho(t)&=&-\frac{i}{2}S(t)
  [\sigma^+\sigma^-,\rho(t)]\\
  &&\hspace*{-3em}+\gamma(t)\left\{-\frac{1}{2}\sigma^+\sigma^-\rho(t)
  -\frac{1}{2}\rho(t)\sigma^+\sigma^- +\sigma^-\rho(t)\sigma^+
  \right\},\nonumber
\end{eqnarray}
where the time-dependent coefficients $S(t)$ and $\gamma(t)$ are given
by 
\begin{equation}
  \label{eq:S_gamm}
  S(t)=-2\Im\left\{\frac{\dot c_1(t)}{c_1(t)}\right\},\quad
  \gamma(t)=-2\Re\left\{\frac{\dot c_1(t)}{c_1(t)}\right\}.
\end{equation}
It is important to note that the time-convolutionless expansion of the
equation of motion (\ref{eq:TCL-app}) reproduces the structure of the
exact equation of motion (\ref{eq:TCL-QME}) to all orders in the
coupling. On the other hand, a perturbative expansion of the exact
Nakajima-Zwanzig equation to fourth order \cite{ShibataJPhysJ49} also
contains terms of the form $\sigma^+\sigma^-\rho(s)\sigma^+\sigma^-$.

In order to find a stochastic unravelling of the time-convolutionless
quantum master equation, we have to distinguish two cases: if the
function $J(\omega)$ leads to a rate $\gamma^{(4)}(t)$ which is
positive for all $t$, then we can use the ``usual'' stochastic
unravelling described in Sec.~\ref{sec:stlind}. Otherwise, if the rate
also takes negative values, we have to use the more general algorithm
which we presented in Sec.~\ref{sec:SGQME}. We will illustrate both
cases by means of the damped Jaynes-Cummings model, which describes
the coupling of a two-level system to a single cavity mode, which in
turn is coupled to an environment. For this model, the function
$J(\omega)$ is given by 
\begin{equation}
  \label{eq:JC}
  J(\omega)=\frac{1}{2\pi}\frac{\gamma_0\lambda^2} 
  {(\omega_0-\omega)^2+\lambda^2},
\end{equation}
where $\omega_0$ is the center frequency of the cavity. Using
Eq.~(\ref{eq:phi_psi}) we obtain
\begin{eqnarray}
  \label{eq:phi}
  \Phi(t)&=&\gamma_0\lambda e^{-\lambda t}\cos(\Delta t),\\
  \label{eq:psi}
  \Psi(t)&=&\gamma_0\lambda e^{-\lambda t}\sin(\Delta t),
\end{eqnarray}
where the detuning $\Delta$ is defined as $\Delta=\omega_{\rm S}-\omega_0$.
Obviously, for $\Delta=0$, the function $\Phi(t)$ is purely
exponential, and $\Psi(t)$ vanishes. Hence, $\gamma^{(4)}$ is
positive for all $t$. On the other hand, for $\Delta\ne 0$ the rate
$\gamma^{(4)}(t)$ oscillates and can take negative values if $\Delta$
is sufficiently large. We will discuss both cases separately.

%----------------------------------------------------------

\subsection{Damped Jaynes-Cummings model on resonance}
\label{sec:JCres}

\begin{figure}[t]
  \begin{center}
    \leavevmode \epsfxsize\linewidth\epsffile{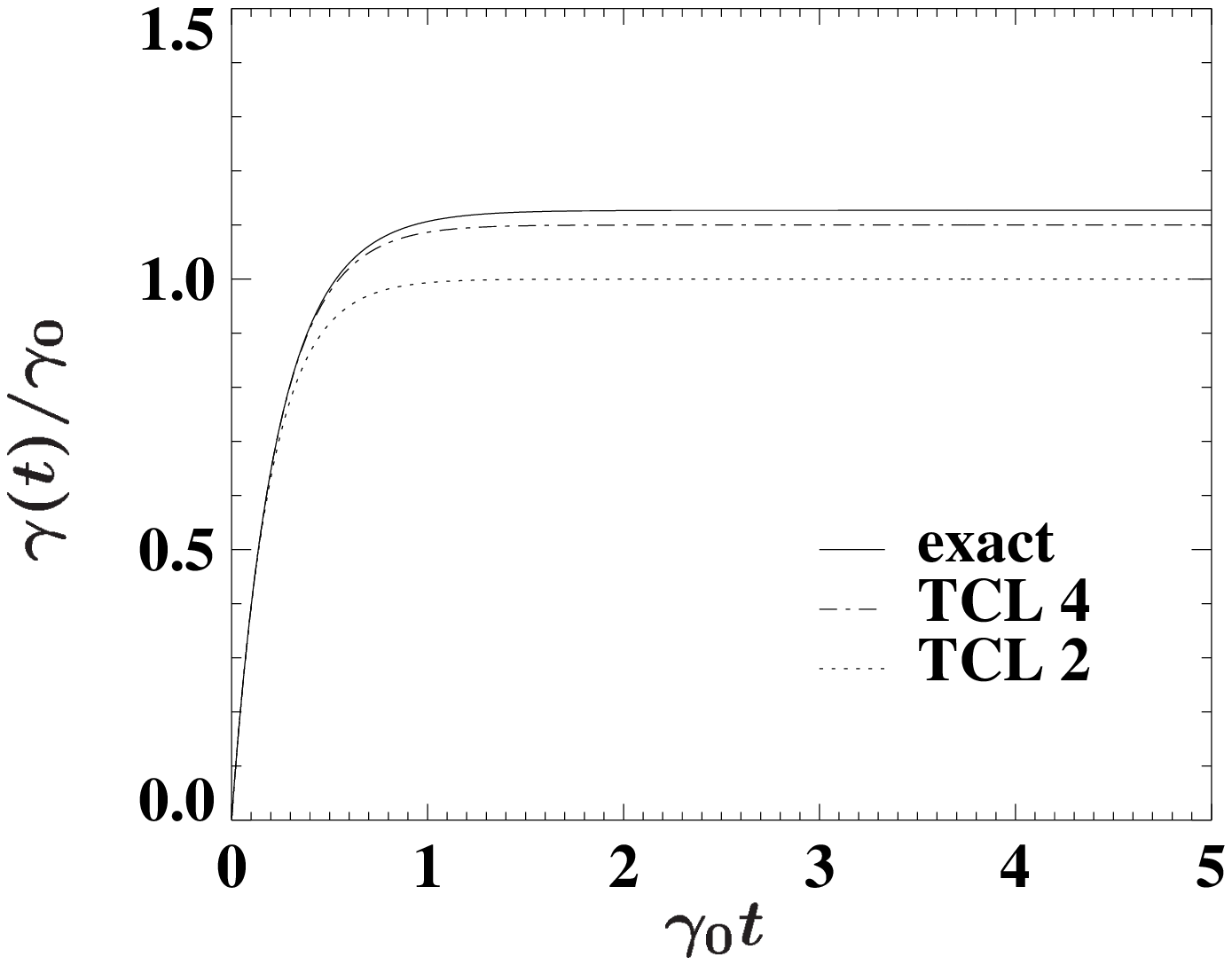}
    \caption{Decay rate of the excited state in the damped 
      Jaynes-Cummings model using the time-convolutionless master equation to
      second (TCL 2) and fourth (TCL 4) order, compared
      with the exact decay rate. The reservoir correlation time is
      $\tau_{\rm R}=0.2\gamma_0^{-1}$. }
  \label{fig:1}
  \end{center}
\end{figure}

For the resonant damped Jaynes-Cummings model, the rate
$\gamma^{(4)}(t)$ can be calculated using Eq.~(\ref{eq:rate}). It is given by 
\begin{equation}
  \label{eq:gamm4TCL}
  \gamma^{(4)}(t)=\gamma_0\left\{1-e^{-\lambda t}+\frac{\gamma_0}{\lambda}
  \left[\sinh(\lambda t)-\lambda t\right]e^{-\lambda t}\right\},
\end{equation}
which we have illustrated in Fig.~\ref{fig:1} for $\lambda=5\gamma_0$
together with the rate $\gamma^{(2)}(t)$ and the exact decay rate
\begin{equation}
  \label{eq:gamma_ex}
  \gamma(t)=\frac{2\gamma_0\lambda\sinh({dt}/{2})}
  {d\cosh({dt}/{2})+\lambda\sinh({dt}/{2})},
\end{equation}
where $d=\sqrt{\lambda^2-2\gamma_0\lambda}$. The exact rate is
obtained by inserting the amplitude $c_1(t)$ (see
\cite{GarrawayPRA55}) into Eq.~(\ref{eq:S_gamm}).

\begin{figure}[t]
  \begin{center}
    \leavevmode \epsfxsize\linewidth\epsffile{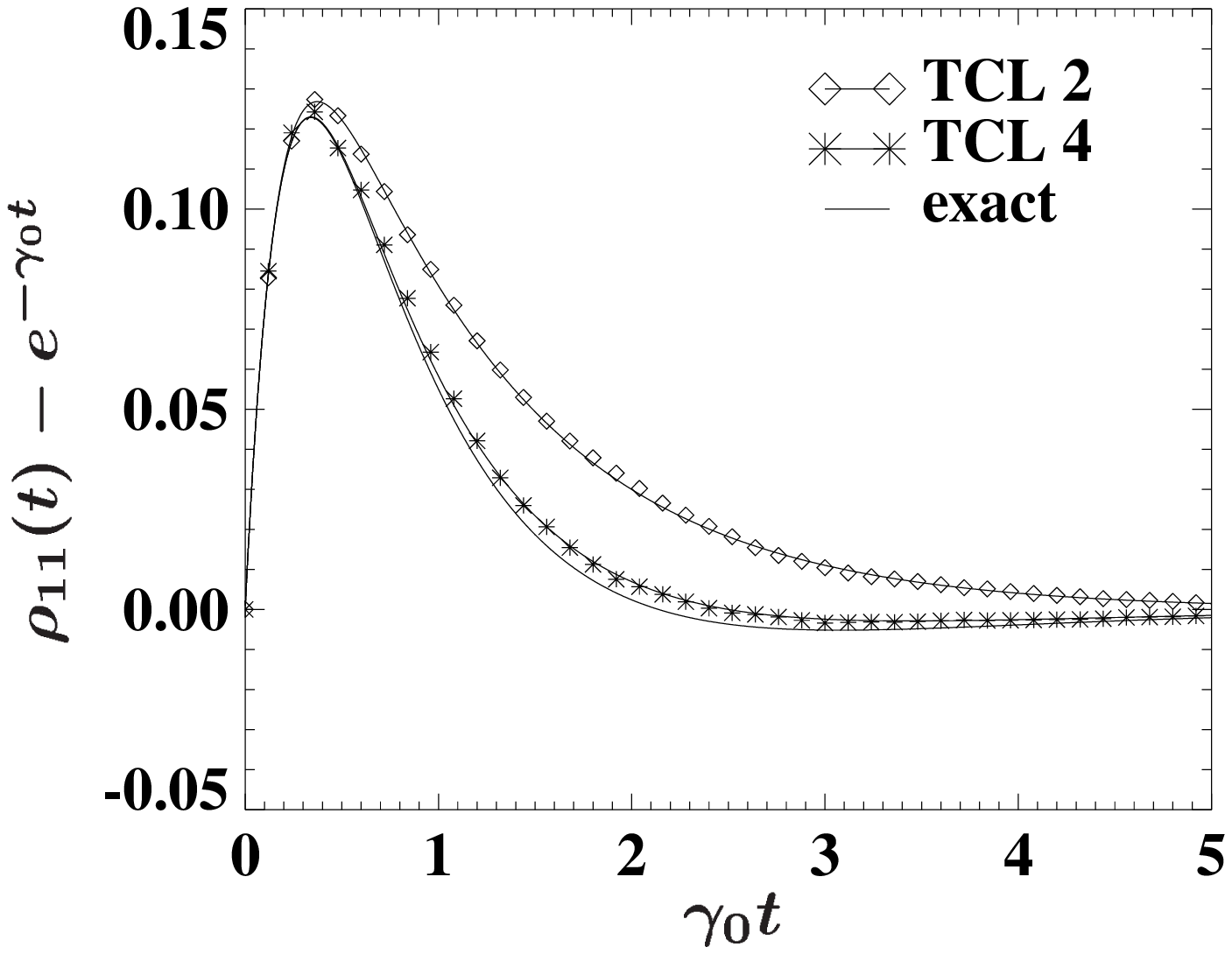}
    \caption{Deviation of the population $\rho_{11}(t)$ from the
      Markovian population $e^{-\gamma_0 t}$ for the damped
      Jaynes-Cummings model for the inverse reservoir correlation time
      $\lambda=5\gamma_0$: stochastic simulation (symbols)
      and exact solution of the time-convolutionless master equation
      to second (TCL 2) and fourth (TCL 4) order, compared with the
      exact solution. }
  \label{fig:2}
  \end{center}
\end{figure}

Since the rate $\gamma^{(4)}(t)$ is positive for all $t$ we may use
the unravelling presented in Sec.~\ref{sec:stlind}. The dynamics of the
stochastic wave function is governed by the stochastic differential
equation (\ref{eq:SDE}), where for this model the generator
$G(\psi,t)$ of the deterministic motion is given by
\begin{equation}
  \label{eq:G_res}
  G(\psi,t)=-\frac{i}{2}\gamma^{(4)}(t)\left(\sigma^+\sigma^-
  -\|\sigma^-\psi\|^2\right)\psi
\end{equation}
and the instantaneous transitions lead to jumps of the form
\begin{equation}
  \label{eq:trans_res}
  \psi(t)\longrightarrow\frac{\sigma^-\psi(t)}{\|\sigma^-\psi(t)\|}= 
  |0\rangle_{\rm S},
\end{equation}
i.~e., the state of the system is projected onto the ground state. In
Fig.~\ref{fig:2} we illustrate the deviation of $\rho_{11}(t)$ from
the Markovian population $\exp(-\gamma_0 t)$ for an initially excited
system. Obviously, the perturbative expansion of the
time-convolutionless generator $K(t)$ converges rapidly and leads
to an excellent agreement of the exact solution and the solution of
the time-convolutionless quantum master equation to fourth order. In
Fig.~\ref{fig:2} we also show the solution of the stochastic
simulation of Eq.~(\ref{eq:SDE}) for $10^5$ realizations, which is in
very good agreement with the solution of the time-convolutionless
quantum master equation (\ref{eq:TCL-app}).

%----------------------------------------------------------

\subsection{Damped Jaynes-Cummings model with detuning}
\label{JCdet}

\begin{figure}[t]
  \begin{center}
    \leavevmode \epsfxsize\linewidth\epsffile{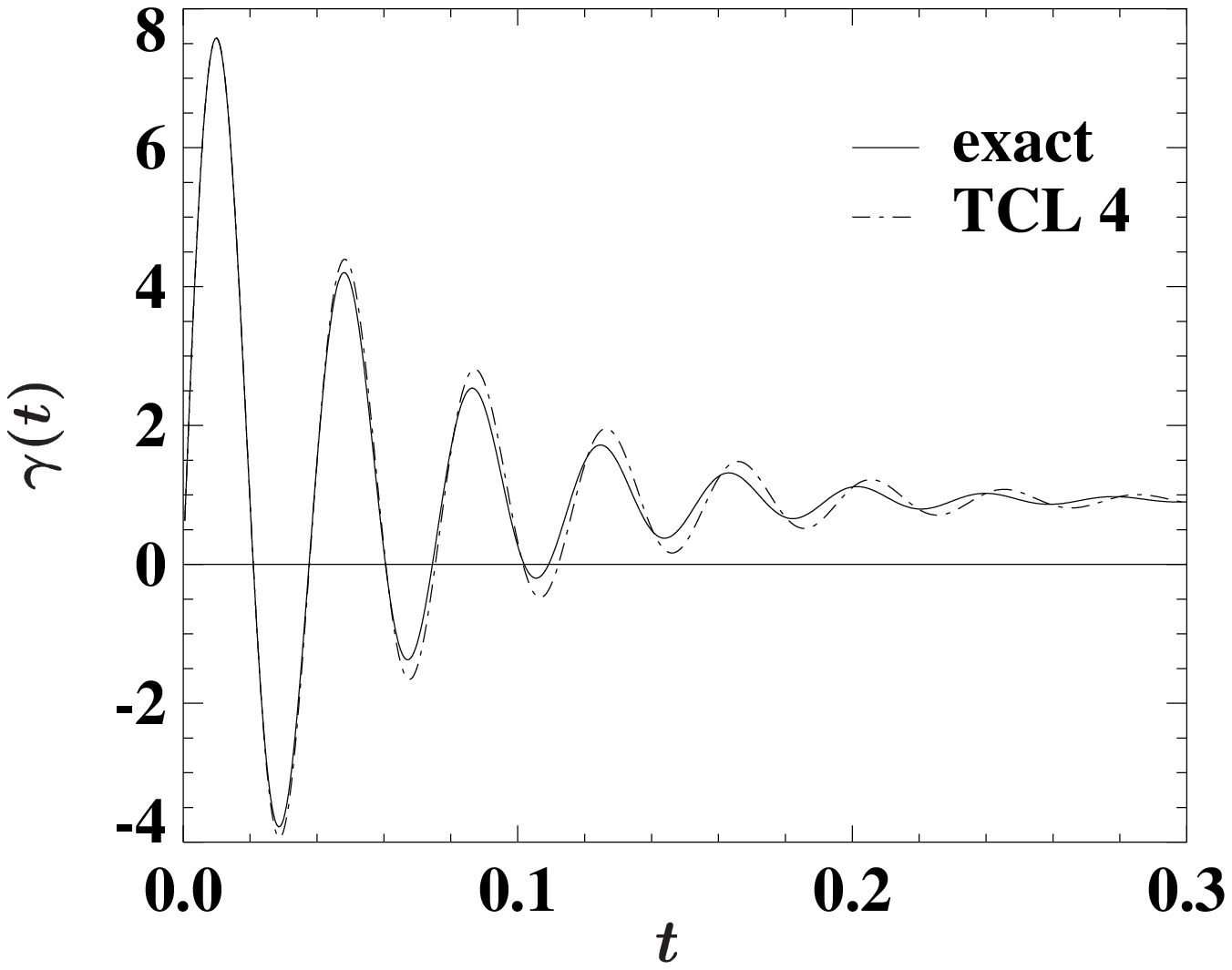}
    \caption{Decay rate of the excited state in the damped 
      Jaynes-Cummings model with detuning using the
      time-convolutionless master equation to 
      fourth order (TCL 4) compared with the exact decay rate. The
      parameters are: $\gamma_0=65$, $\lambda=19.5$, and $\Delta=8\lambda$.}
  \label{fig:3}
  \end{center}
\end{figure}

Making use of Eq.~(\ref{eq:rate}) we can calculate the decay rate
$\gamma^{(4)}(t)$ for the damped Jaynes-Cummings model with detuning,
which yields
\begin{eqnarray}
  \label{eq:rate_det}
  \lefteqn{\gamma^{(4)}(t)=\frac {\gamma_0\lambda^2 }{\lambda ^{2}+\Delta ^{2}}
  \left[1-{e^{-\lambda t}}\left(\cos(\Delta t)
  -{\textstyle\frac{\Delta}{\lambda}}\sin(\Delta t)\right)\right]}\nonumber\\
 && +\frac{\gamma_0^2\lambda^5e^{-\lambda t}}{2(\lambda^2+\Delta^2)^3}\Big\{
   \Big[1-3\left({\textstyle\frac{\Delta}{\lambda}}\right)^2\Big]\left(e^{\lambda
  t}-e^{-\lambda t}\cos(2\Delta t)\right)\nonumber\\
  && -2\Big[1-\left({\textstyle\frac{\Delta}{\lambda}}\right)^4\Big]\lambda
  t\cos(\Delta t)
   +4\Big[1+\left({\textstyle\frac{\Delta}{\lambda}}\right)^2\Big]\Delta
  t\sin(\Delta t)\nonumber\\
  &&+{\textstyle\frac{\Delta}{\lambda}}\Big[3-\left({\textstyle\frac{\Delta}{\lambda}}\right)^2\Big] 
  e^{-\lambda t}\sin(2\Delta t) 
  \Big\}.
\end{eqnarray}
In Fig.~\ref{fig:3} we have depicted the rate $\gamma^{(4)}(t)$
together with the exact decay rate. The parameters are chosen
such that performing the usual Born-Markov approximation leads to
the constant decay rate $\gamma_{\rm M}=1$. For short times, the decay rate
shows damped oscillations and converges in the long time limit to a
time-independent decay rate which is close to $\gamma_{\rm M}$.

\begin{figure}[t]
  \begin{center}
    \leavevmode \epsfxsize\linewidth\epsffile{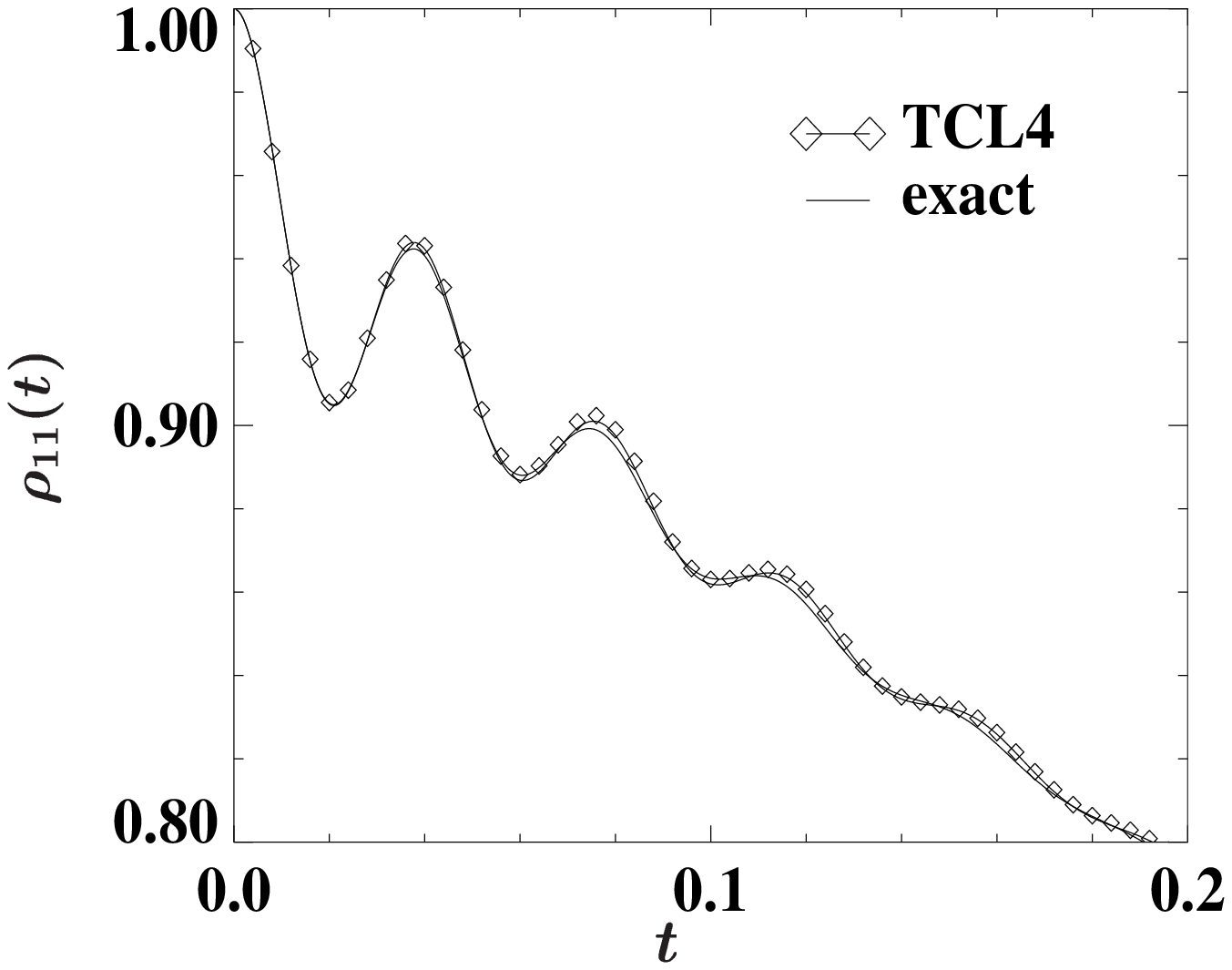}
    \caption{Decay of the population $\rho_{11}(t)$ for the  damped
      Jaynes-Cummings model with detuning using the
      time-con\-vo\-lu\-tion\-less master equation to fourth order (TCL 4)
      compared with the exact decay rate. The parameters are:
      $\gamma_0=65$, $\lambda=19.5$, and $\Delta=8\lambda$}
  \label{fig:4}
  \end{center}
\end{figure}

Note however, that in this case the decay rate can also take negative values,
and the corresponding quantum master equation has to be unraveled
using the procedure described in Sec.~\ref{sec:SGQME}. Thus, the
dynamics of the stochastic wave function
$\theta(t)=(\phi(t),\psi(t))^T$, being an element of the doubled
Hilbert space $\widetilde{\cal H}={\cal H}\oplus{\cal H}$ is governed by the 
stochastic differential equation (\ref{eq:gen_un}), where the
operators $F$ and $J$ are given by
\begin{equation}
  \label{eq:F_dbl}
  F=-\frac{1}{2}\gamma^{(4)}(t)\left(\begin{array}{cc}
   \sigma^+\sigma^- &0\\
  0&\sigma^+\sigma^- 
  \end{array}\right)
\end{equation}
and
\begin{equation}
  \label{eq:J_dbl}
  J=\left(\begin{array}{cc}
  \gamma^{(4)}(t)\sigma^-&0\\
  0& \sigma^-
  \end{array}\right).
\end{equation}
Here, the generator of the deterministic motion is given by
\begin{equation}
  \label{eq:G_det}
   G(\theta,t)=i\left( F+\frac{1}{2}
  \frac{\left\| J\theta(t)\right\|^2}{\|\theta(t)\|^2}\right)
  \theta(t).
\end{equation}
and the jumps induce instantaneous transitions of the form
\begin{equation}
  \label{eq:dbl_jump}
  \theta(t)\longrightarrow\frac{\|\theta(t)\|}{\|J\theta(t)\|}J\theta(t)
  \sim\left(
    \begin{array}{c}
      \gamma^{(4)}(t)|0\rangle_{\rm S}\\
      |0\rangle_{\rm S}
    \end{array}
\right).
\end{equation}
Hence, if the rate $\gamma^{(4)}(t)$ is positive, the jump leads to a
positive contribution to the ground state population $\rho_{00}(t)$,
whereas a negative rate leads to a negative contribution to
$\rho_{00}$.  The results of a stochastic simulation of
Eq.~(\ref{eq:gen_un}) with $10^5$ realizations is displayed in
Fig.~\ref{fig:4} together with the analytical solution of the
time-convolutionless quantum master equation to fourth order
(\ref{eq:TCL-app}) and the exact solution, which are in very good
agreement. This clearly demonstrates the usefulness of the simulation
algorithm presented here.

%----------------------------------------------------------
%   Conclusion
%----------------------------------------------------------
\section{Summary}
\label{sec:Summ}

In this article we have presented a generalization of the stochastic wave
function method to arbitrary linear quantum master equations, which
allows an explicit treatment of memory effects in a systematic
way. This is done by employing the time-convolutionless projection
operator technique, which yields a perturbative expansion of the
equation of motion of the reduced density matrix. The latter is then
unraveled by a stochastic wave function in the doubled Hilbert space.
By means of the damped Jaynes-Cummings model, which is an exactly
solvable model, we have illustrated the general theory and tested the
performance of this method.

\section*{Acknowledgment}

HPB would like to thank the Istituto Italiano per gli Studi Filosofici
in Naples (Italy) and BK would like to thank the DFG-Graduiertenkolleg
{\it Nichtlineare Differentialgleichungen} at the
Albert-Ludwigs-Universit\"at Freiburg for financial support of the
research project.

% --------- Bibliography ------------------------

\bibliographystyle{prsty}  % Phys. Rev. / APS Style

\newpage

%----------Fig. 1------------------------------

%----------Fig. 2------------------------------

%----------Fig. 3------------------------------

%----------Fig. 4------------------------------
\end{multicols}
\end{document}